\documentclass[twocolumn,prl,amsmath,amssymb,showpacs,superscriptaddress,floatfix]{revtex4}
\usepackage{graphicx}
\usepackage{bm}

\usepackage{epsf}
\begin{document}
\title{Random pinning elucidates the nature of melting transition in
two-dimensional core-softened potential system}

\author{E. N. Tsiok}
\affiliation{ Institute for High Pressure Physics RAS, 108840
Kaluzhskoe shosse, 14, Troitsk, Moscow, Russia}

\author{Yu. D. Fomin}
\affiliation{ Institute for High Pressure Physics RAS, 108840
Kaluzhskoe shosse, 14, Troitsk, Moscow, Russia}

\author{V. N. Ryzhov}
\affiliation{ Institute for High Pressure Physics RAS, 108840
Kaluzhskoe shosse, 14, Troitsk, Moscow, Russia}

\date{\today}

\begin{abstract}
Despite about forty years of investigations, the nature of the
melting transition in two dimensions is not completely clear. In
the framework of the most popular
Berezinskii-Kosterlitz-Thouless-Halperin-Nelson-Young (BKTHNY)
theory, 2D systems melt through two continuous
Berezinskii-Kosterlitz-Thouless (BKT) transitions with
intermediate hexatic phase. The conventional first-order
transition is also possible. On the other hand, recently on the
basis of computer simulations the new melting scenario was
proposed with continuous BKT type solid-hexatic transition and
first order hexatic-liquid transition. However, in the simulations
the hexatic phase is extremely narrow that makes its study
difficult. In the present paper, we propose to apply the random
pinning to investigate the hexatic phase in more detail. The
results of molecular dynamics simulations of two dimensional
system having core-softened potentials with narrow repulsive step
which is similar to the soft disk system are outlined. The system
has a small fraction of pinned particles giving quenched disorder.
Random pinning widens the hexatic phase without changing the
melting scenario and gives the possibility to study the behavior
of the diffusivity and order parameters in the vicinity of the
melting transition and inside the hexatic phase.
\end{abstract}


\maketitle

\section{Introduction}

The phenomenon of two-dimensional melting is one of the
long-standing issues addressed by condensed matter physics.
Despite numerous papers publisher over the last forty years, the
mechanisms underlying the microscopic nature of melting in two
dimensions ($2D$)  remain unclear. In contrast to the three
dimensional case, where melting develops through conventional
first order transition, several microscopic scenarios have been
proposed to explain microscopic $2D$ melting. The difference
between these cases deals with dramatic increase of fluctuations
in $2D$ in comparison with the $3D$ case. Landau, Peierls and
Mermin \cite{landau,mermin} showed that in two dimension situation
the long-ranged positional order can not exist because of the
thermal fluctuations, therefore the  positional order transforms
into the quasi-long-ranged order. On the other hand, the real
long-ranged orientational order (the order in directions of the
bonds binding a particle with its neighbors) does exist in this
case.

Currently, the
Berezinskii-Kosterlitz-Thouless-Halperin-Nelson-Young (BKTHNY)
theory of $2D$ melting
\cite{ber1,ber2,kosthoul72,kosthoul73,halpnel1,halpnel2,halpnel3}
is considered the best to derscribe the phenomenon. According to
this theory $2D$ solids melt through two continuous transitions,
which are activated by topological defects. For example, a
disclination is an elementary topological defect in triangular
lattice. In a triangular crystal lattice disclination is defined
as an isolated defect having five or seven neighbours. A
dislocation can be considered as a bound pair of 5- and 7-fold
disclinations. In the framework of BKTHNY scenario, $2D$ solids
melt through dissociation of bound dislocation pairs. In this case
the long-ranged orientational order transforms into
quasi-long-ranged order, and the quasi-long-ranged positional
order becomes short-ranged. The new intermediate phase is called a
hexatic phase. In turn, the hexatic phase transforms into an
isotropic liquid phase having short-ranged orientational and
positional orders through unbinding dislocations (5 and 7-fold
bound pairs) into free disclinations. It should be noted, that the
BKTHNY theory provides only limits of stability of the solid and
hexatic phases and does not rule out that the first  order
liquid-solid transition can be preceded by other melting
mechanisms.

The BKTHNY theory seems universal and it can applied to describe
all systems. It does not contain the interparticle potential in
the explicit form, but, it contains two phenomenological
parameters - the core energy of dislocation $E_c$ and Frank module
of hexatic phase $K_A$ - that cannot be expressed explicitly in
terms of the interparticle potential. Later it was shown that
decreasing $E_c$ can result in melting occuring through a single
first-order transition resulting, for example, from forming grain
boundaries \cite{chui83} or "exploding" 5-7-5-7 quartets (bound
dislocation pairs) into free disclinations \cite{ryzhovJETP}.

Recently the BKTHNY scenario was confirmed in experiments with
superparamagnetic colloidal particles interacting through
long-range dipolar potential \cite{keim1,zanh,keim2,keim3,keim4}.
In these experiments the particles were absorbed at a liquid-air
interface which restricted the out-of-plane motion. On the other
hand, melting transition consistent with the BKTHNY scenario was
found \cite{col2,col3,col4} in popular experimental systems of
colloidal particles confined between two glass plates \cite{col1}.
At the same time, the exsistance of the first order liquid-solid
transitions \cite{col5} and even the first-order liquid-hexatic
and the first-order hexatic-solid phase transitions \cite{col6}
can be expected. In 2D superconducting vortex lattices,
macroscopic measurements provide evidence for melting close to the
transition to the normal state. In Refs. \cite{sup1,sup2} the
authors using the scanning tunnelling spectroscopy showed directly
that the transition into an isotropic vortex liquid below the
superconducting critical temperature does exists. Before that,
they found a hexatic phase, characterized by the appearance of
free dislocations, and a smectic-like phase, possibly formed
through partial disclination unbinding. These results confirm that
the melting mechanism is not universal and it depends on
interparticle interactions. At the same time the ambiguity remains
even in describing the same systems such as, for example, the
systems of hard disks
\cite{rto1,rto2,RT1,RT2,DF,strandburg92,binderPRB,mak,binder,LJ}.

Recently, the other melting scenario was proposed
\cite{foh1,foh2,foh3,foh4,foh5,foh6}. In contrast to the BKTHNY
theory, it was maintained that the hexatic phase does exist in the
basic hard disk model, and the system melts through  a continuous
solid-hexatic transition but through a first-order hexatic-liquid
transition \cite{foh1,foh2,foh3}. In \cite{foh5} it was shown that
a first-order transition occurs between the stable hexatic phase
and isotropic liquid in $2D$ Yukawa system. In paper \cite{foh4}
Kapfer and Krauth explored behavior of a soft disk system with
potential
 $U(r)=(\sigma/r)^n$. The system was shown to
melt in accordance with the BKTHNY theory for $n\leq 6$, while for
$n>6$ the two-stage melting transition took place with continuous
solid-hexatic and the first-order hexatic-liquid transition. This
scenario was confirmed experimentally in a system of colloidal
particles on water-decane interface \cite{col7}. Recently a paper
was published \cite{foh7} where at low densities the Herzian disks
model was shown to demonstrate the reentering melting transition
featuring the maximum on the melting line. The discontinuous
liquid-hexatic transition occurs at lower than the maximum
densities, while at higher densities systems melt undergoing a
continuous BKT transition.

It should be noted that the simulations in Refs.
\cite{foh1,foh2,foh3,foh4,foh5,foh6} demonstrate the very narrow
hexatic phase. However, the presence of disorder can widen the
hexatic phase and help study its properties. Experiments usually
involve two-dimensional confinement provided by slit pores having
different origin or by adsorption on solid substrates. Both cases
can result in the frozen-in (quenched) disorder generated by a
certain roughness. Quenched disorder can change the melting
scenario in $2D$. A disordered substrate  can have a similar
destructive effect on the crystalline order as temperature and can
bring about melting even at zero temperature
\cite{nel_dis1,nel_dis2,dis3,dis4}. It was shown in Refs.
\cite{nel_dis1,nel_dis2} that the BKTHNY melting scenario persists
in presence of weak disorder. It is intuitively understood that
the temperature of the hexatic-isotropic liquid transition $T_i$
is almost unaffected by disorder, whereas the melting temperature
$T_m$ decreases drastically with increasing disorder
\cite{dis3,dis4,nel_dis1,nel_dis2}. As a result, the stability
range of the hexatic phase gets wider. These predictions were
confirmed in experiment and through simulating superparamagnetic
colloidal particle systems \cite{keim3,keim4}. In these
experiments under gravity particles formed a monolayer at the
bottom of a cylindrical glass cell having $5-mm$ diameter.
Quenched disorder developed due to pinning a small amount of
particles to the glass substrate due to van der Waals interactions
and chemical reactions. In our simulations we tried to find a
simulation method similar to above experimental setup.

In this paper we present results of detailed computer  simulation
studies of $2D$ phase diagram previously suggested core softened
potential system \cite{jcp2008,wepre,we_inv,we2011,RCR,we2013-2}
in presence of quenched disorder for small value of the width of
repulsive shoulder. The different forms of core-softened
potentials are widely used for qualitative description of the
anomalous water-like behavior, including density, structural and
diffusion anomalies, liquid-liquid phase transitions, glass
transitions, and melting maxima
\cite{jcp2008,wepre,we_inv,we2011,RCR,we2013-2,buld2009,fr1,fr2,fr3,fr4,
barbosa,barbosa1,barbosa2,barbosa3,buld2d,scala,prest2,prest1}. In
our previous publications the preliminary results on plotting
phase diagrams for the system were reported
\cite{dfrt1,dfrt2,dfrt3,dfrt4,dfrt5}.

Here we approached the case of the system having small repulsive
shoulder. It was shown that in this case the behavior of the
system potential was similar to that of a soft disks system
\cite{dfrt2}. The random pinning results in drastic increase of
the width of hexatic  phase without changing the melting scenario.
It gives the possibility to study some properties of the hexatic
phase including calculation of the diffusion coefficient and the
behavior of the orientational and translational order parameters.

\section{Systems and methods}

\begin{figure}
\begin{center}
\includegraphics[width=8cm]{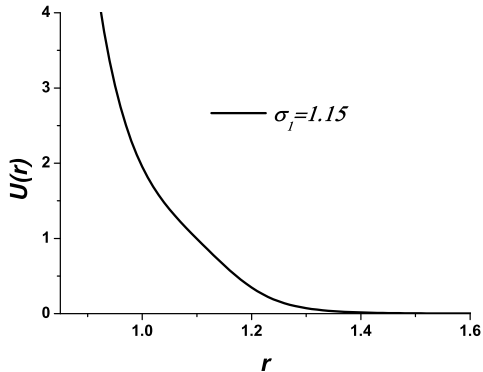}%

\end{center}

\caption{\label{fig:fig1} The potential (\ref{3}) with soft-core
diameter $\sigma_1/\sigma = 1.15$.}
\end{figure}

In the current simulations we studied a system described
by potential
\cite{jcp2008,wepre,we_inv,we2011,RCR,we2013-2,dfrt1,dfrt2,dfrt3,dfrt5}:
\begin{equation}
U(r)=\varepsilon\left(\frac{\sigma}{r}\right)^{n}+\frac{1}{2}\varepsilon\left(1-
\tanh(k_1\{r-\sigma_1\})\right). \label{3}
\end{equation}
where $n = 14$ and $k_1\sigma = 10.0$. $\sigma$ are the hard-core
diameters. We simulate systems having small soft-core diameter:
$\sigma_1/\sigma = 1.15$ (see Fig.~\ref{fig:fig1}).

Through the paper we used dimensionless quantities, which in $2D$
had the form: $\tilde{{\bf r}}\equiv {\bf r}/\sigma$,
$\tilde{P}\equiv P \sigma^{2}/\varepsilon ,$ $\tilde{V}\equiv V/N
\sigma^{2}\equiv 1/\tilde{\rho}, \tilde{T} \equiv
k_BT/\varepsilon, \tilde{\sigma}=\sigma_1/\sigma$.  Hereinafter we
omitted the tildes.

The study dealt with molecular dynamics simulations  in $NVT$ and
$NVE$ ensembles in the framework of the LAMMPS package
\cite{lammps} for the number of particles ranging from $20000$ to
$100000$. We randomly choose a subset of particles at the random
positions and kept them immobile for a complete simulation run
\cite{dfrt5} in order to model a quenched disorder. The
simulations of $10$ independent replicas of the system with
different distributions of random pinned patterns were made. The
thermodynamic functions were calculated by averaging over
replicas. We obtained curves for pressure $P$ plotted as functions
of density $\rho$ along isotherms; we plotted correlation
functions $G_6(r)$ and $G_T(r)$ of the bond orientational $\psi_6$
and translational $\psi_T$ order parameters (OPs), which
characterized the overall orientational and translational order
\cite{dfrt5}.

The translational $\psi_T$ (TOP) and orientational order
parameters $\psi_6$ (OOP) and the bond-orientational
$G_6(r)$ (OCF) and translational $G_T(r)$ (TCF) correlation
functions were determined in the ordinary manner
\cite{prest2,halpnel1,halpnel2,binder,binderPRB,LJ,prest1,foh7}
with subsequent averaging over the quenched disorder
\cite{dfrt5}.

In accordance with the standard definitions \cite{halpnel1,
halpnel2, foh7}, TOP has the form:
\begin{equation}
\psi_T=\frac{1}{N}\left<\left<\left|\sum_i e^{i{\bf G
r}_i}\right|\right>\right>_{rp}, \label{psit}
\end{equation}
where ${\bf r}_i$ is the position vector of particle $i$ and {\bf
G} is the reciprocal-lattice vector of the first shell of the
crystal lattice. The translational correlation function can be
obtained from the equation:
\begin{equation}
G_T(r)=\left<\frac{<\exp(i{\bf G}({\bf r}_i-{\bf
r}_j))>}{g(r)}\right>_{rp}, \label{GT}
\end{equation}
where $r=|{\bf r}_i-{\bf r}_j|$ and $g(r)=<\delta({\bf
r}_i)\delta({\bf r}_j)>$  is the pair distribution function. The
second angular brackets $<...>_{rp}$ correspond to averaging
over the random pinning. In the solid phase without random pinning
the long range behavior of $G_T(r)$ has the form $G_T(r)\propto
r^{-\eta_T}$ with $\eta_T \leq \frac{1}{3}$ \cite{halpnel1,
halpnel2}.

To measure the orientational order and the hexatic phase, the
local order parameter determining the $6$-fold orientational
ordering can be defined as follows:
\begin{equation}
\Psi_6({\bf r_i})=\frac{1}{n(i)}\sum_{j=1}^{n(i)} e^{i
n\theta_{ij}}\label{psi6loc},
\end{equation}
where $\theta_{ij}$ is the angle of the vector between particles
$i$ and $j$ with respect to the reference axis and the sum over
$j$ is counting $n(i)$ neighbors of $j$, obtained from the Voronoi
construction. The global OOP can be calculated as an average over
all particles and random pinning:
\begin{equation}
\psi_6=\frac{1}{N}\left<\left<\left|\sum_i \Psi_6({\bf
r}_i)\right|\right>\right>_{rp}.\label{psi6}
\end{equation}

The orientational correlation function $G_6(r)$  (OCF) is provided
in a similar manner in Eq. (\ref{GT}):
\begin{equation}
G_6(r)=\left<\frac{\left<\Psi_6({\bf r})\Psi_6^*({\bf
0})\right>}{g(r)}\right>_{rp}, \label{g6}
\end{equation}
where $\Psi_6({\bf r})$ is the local bond-orientational order
parameter (\ref{psi6loc}). In hexatic phase there is a
quasi-long range order with the algebraic decay of the
orientational correlation function $G_6(r) \propto r^{-\eta_6}$
with $0\leq \eta_6 \leq \frac{1}{4}$
\cite{halpnel1,halpnel2,halpnel3}. The stability criterion of the
hexatic phase has the form $\eta_6(T_i) = \frac{1}{4}$.

The influence of disorder on the orientational and translational
correlation functions was explored in our previous publication
\cite{dfrt5} (see Fig. 1 in \cite{dfrt5}) for $\sigma=1.35$. We
expected that in presence of pinning there would by no qualitative
change observed in the behavior of $G_6(r)$. This was in
accordance with the results by Nelson and coworkers
\cite{nel_dis1, nel_dis2}. On the other hand, the translational
correlation function $G_T(r)$ behaves differently in presence of
random disorder. Without pinning TCF features a conventional power
decay. In case of pinning the envelope gets steeper with crossover
value equal to $r_0$. The region $r<r_0$ corresponds to the local
order unaffected by quenched disorder, whereas asymptotic TCF
behavior when $r>r_0$ is controlled by random pinning
\cite{dfrt5}. It was also shown that according to intuitive
physics expectations  $r_0$ decreases with increase in pinning
centers concentration accompanied by increasing slope of envelope.
The equality $\eta_T(T_m)=1/3$ holding for the long range
asymptote of TCF (for $r>r_0$) can be considered as the
solid-hexatic stability criterion. The hexatic-liquid stability
point corresponds to $\eta_6(T_i)=1/4$ \cite{nel_dis1,nel_dis2}.

\begin{figure}

\includegraphics[width=8cm]{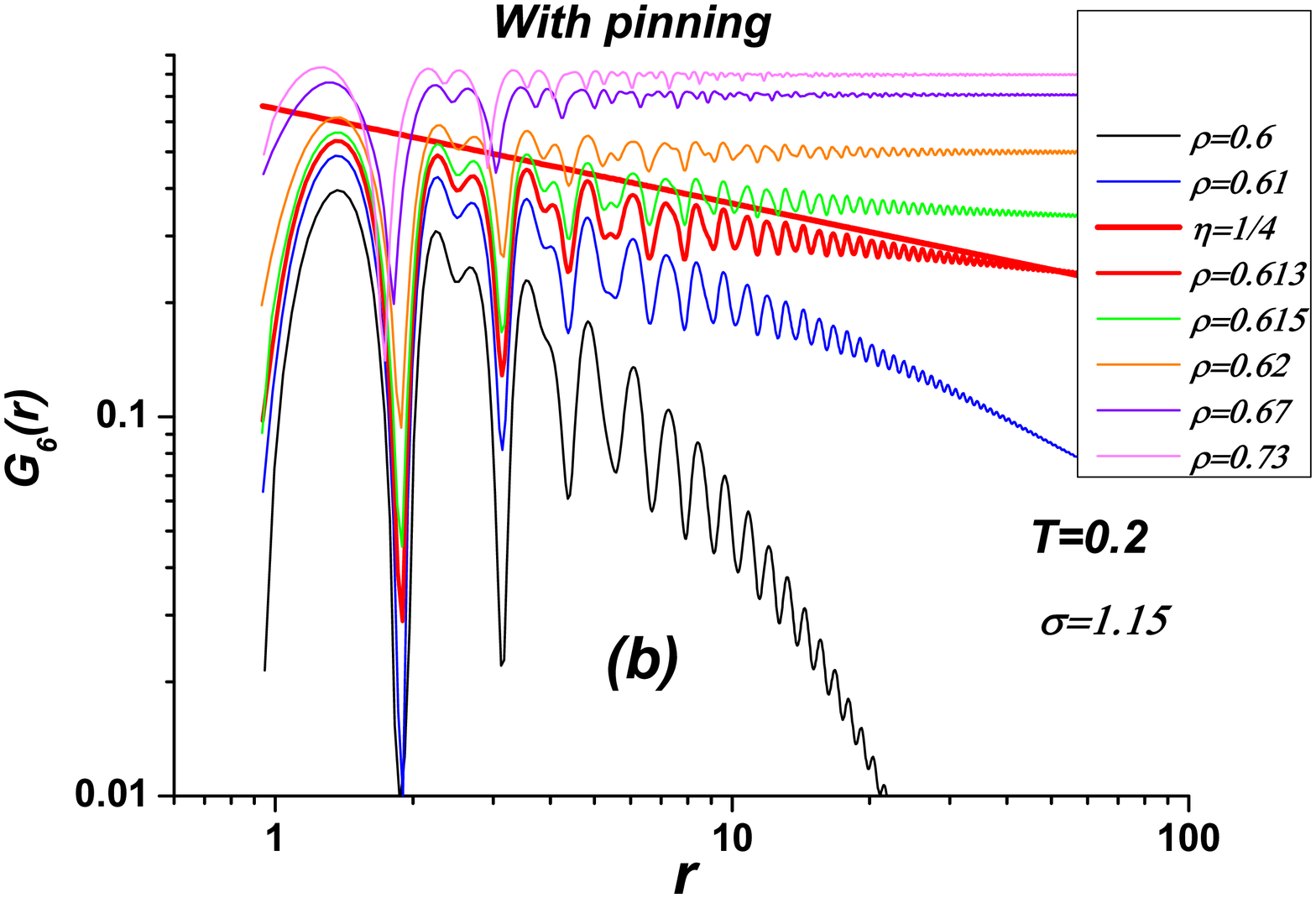}%

\includegraphics[width=8cm]{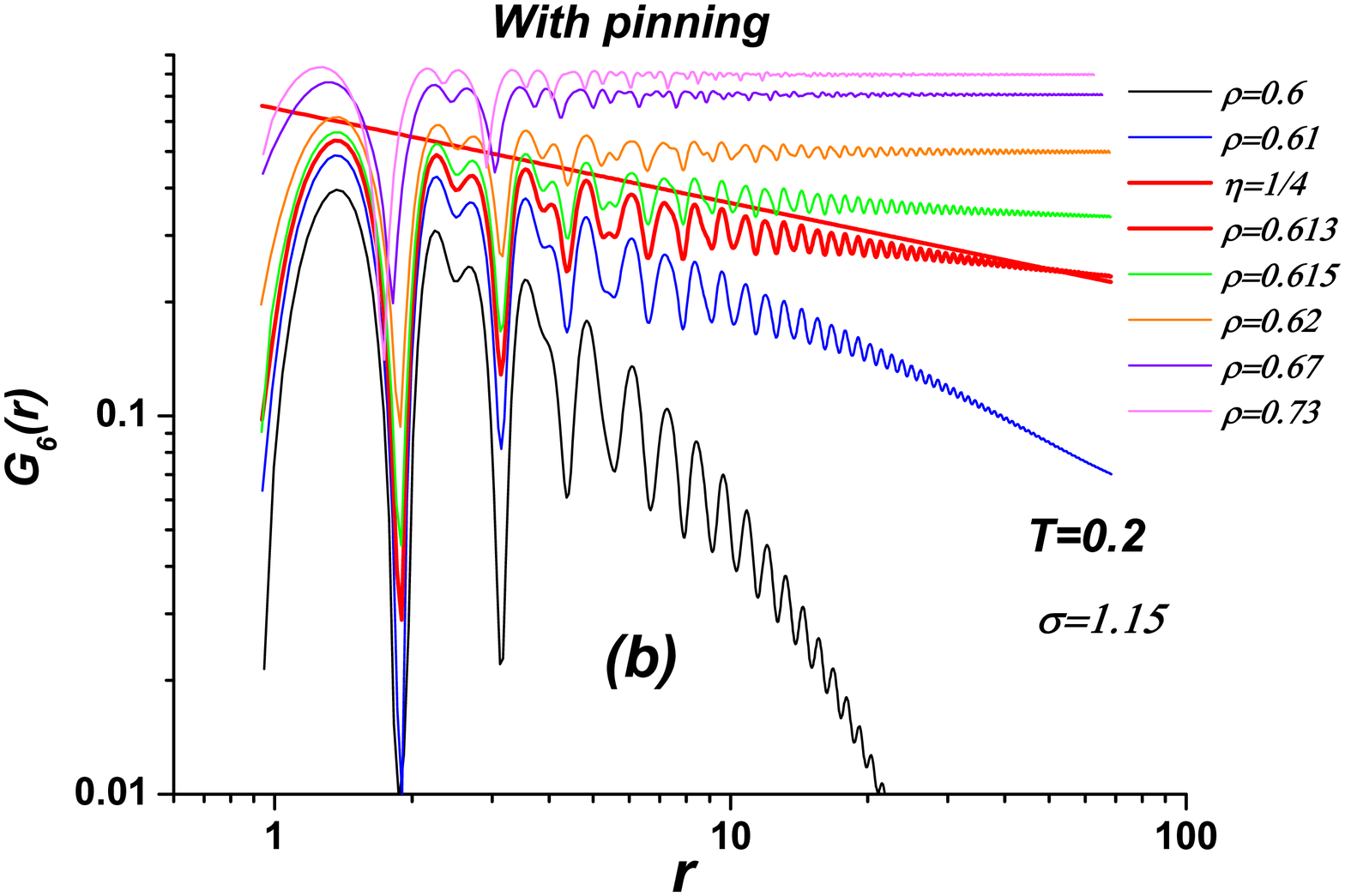}%

\caption{\label{fig:fig2} OCF  plotted for different densities at $T=0.2$
without random pinning (a) and concentration of the pinning
centers $0.1\%$ (b).}
\end{figure}

\section{Results and discussion}

Let us consider the behavior of the system having small repulsive
shoulder $\sigma=1.15$. In this case, as one can see in
Fig.~\ref{fig:fig1}, the form of the potential is similar to that
of a soft disks system with $n=14$. The preliminary discussion of
the phase diagram of this system can be found in Ref.
\cite{dfrt2}, where isotherms with the  van der Waals loops were
developed and the phase diagram was calculated using
double-tangent construction to the Helmholtz free energy curves
\cite{book_fs}.

In Fig.~\ref{fig:fig2} we show the orientational correlation
functions (OCF) obtained for a system without quenched disorder
(a) and having random pinning (concentration of pinning centers is
$0.1\%$) (b) at temperature $T=0.2$. Behavior of OCFs is identical
for both cases, in line with discussions in Introduction section;
the limits of hexatic phase stability determined from condition
$\eta_6=1/4$ also coincide. As Fig.~\ref{fig:fig2} suggests the
density at which hexatic phase becomes unstable is $\rho\approx
0.613$.

\begin{figure}%

\includegraphics[width=8cm]{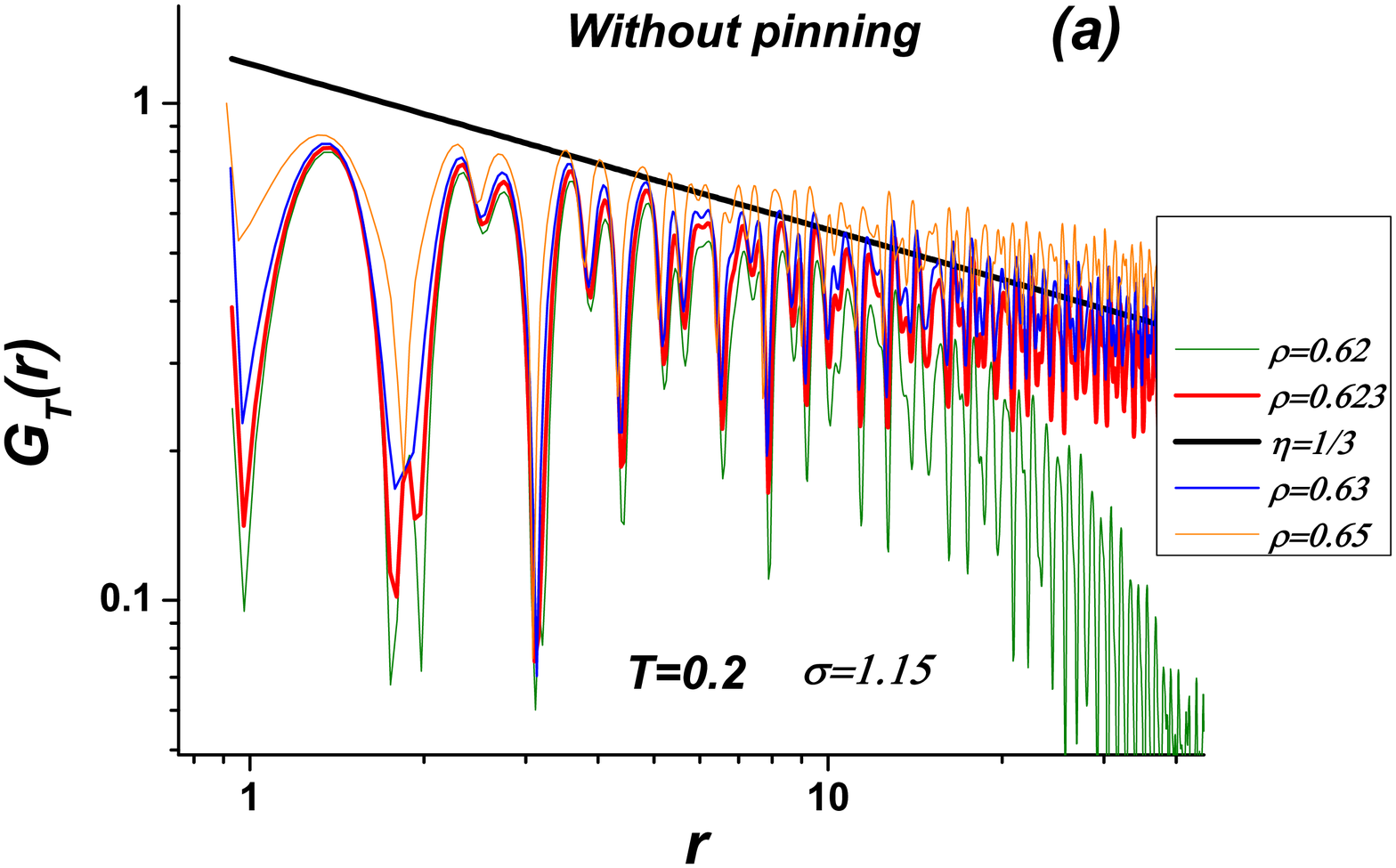}%

\includegraphics[width=8cm]{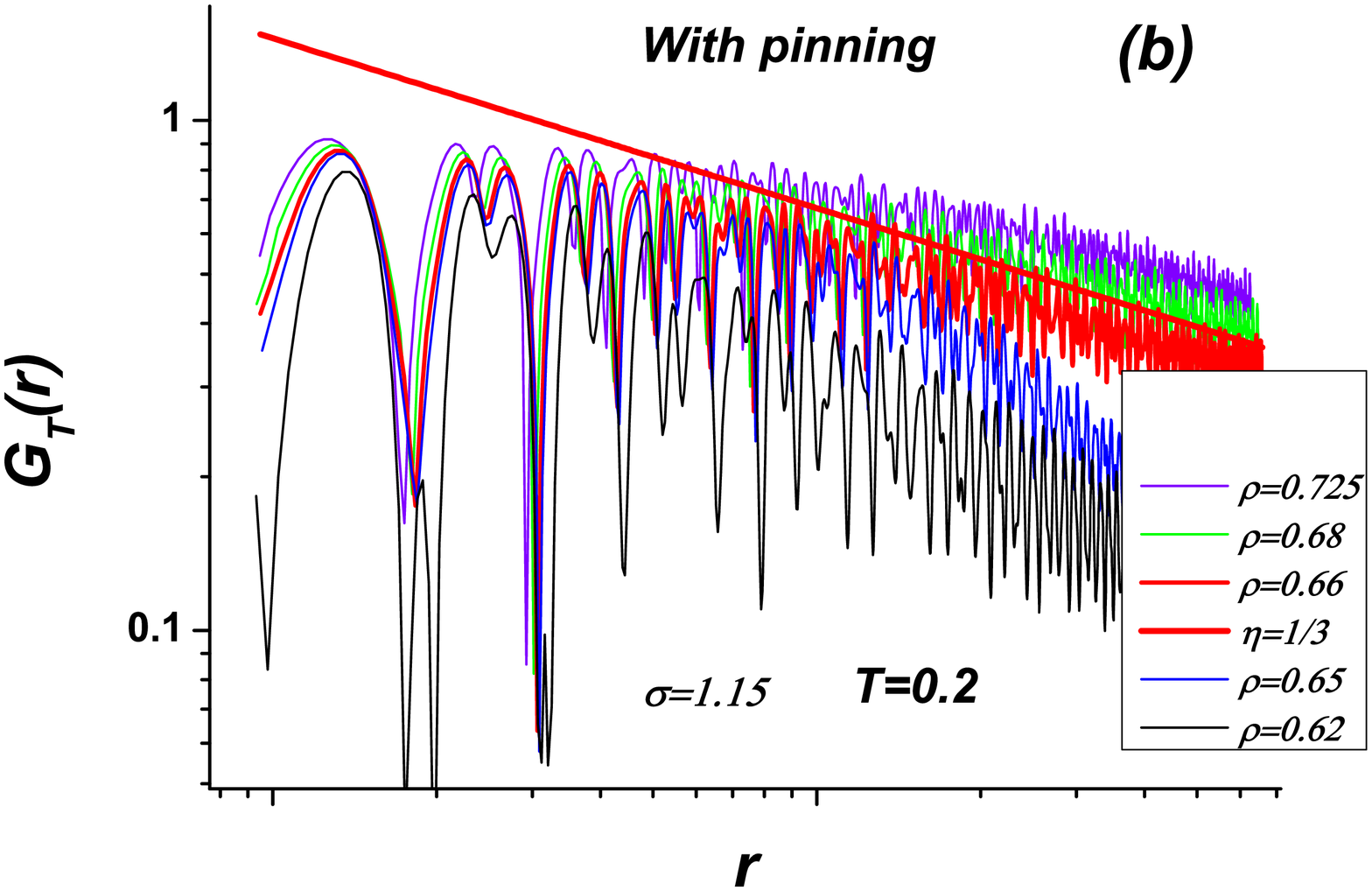}%

\caption{\label{fig:fig3}  TCF  plotted for different densities at $T=0.2$
without random pinning (a) and for concentration of the pinning
centers equal to $0.1\%$ (b).}
\end{figure}

\begin{figure}%

\includegraphics[width=8cm]{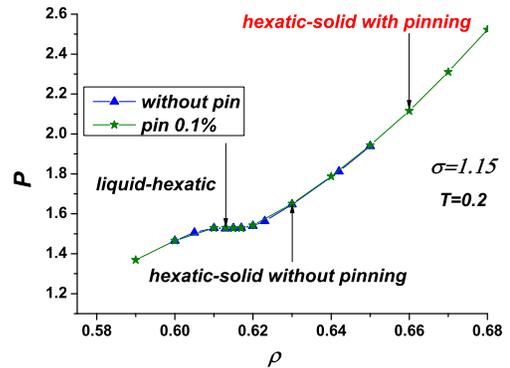}%

\caption{\label{fig:fig4} Isotherms developed for a system having
$\sigma=1.15$ without pinning (triangles) and for system having
concentration of the pinning centers equal to $0.1\%$ (stars) when
$T=0.2$.}
\end{figure}

\begin{figure}
\begin{center}
\includegraphics[width=8cm]{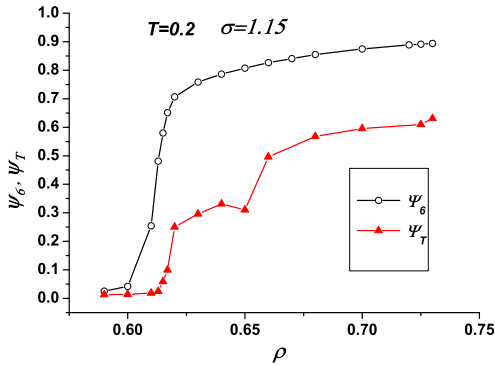}%

\end{center}

\caption{\label{fig:fig5}  The orientational $\Psi_6$ and
translational $\Psi_T$ order parameters for a system having random
pinning and $\sigma = 1.15$.}
\end{figure}

\begin{figure}
\begin{center}
\includegraphics[width=8cm]{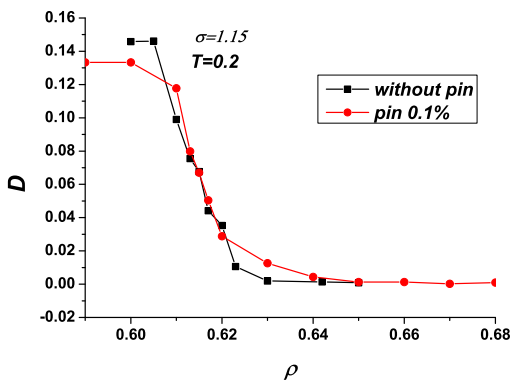}%

\end{center}

\caption{\label{fig:fig6}  The diffusion coefficient of the system
with $\sigma = 1.15$ with (red circles) and without (black
squares) random pinning.}
\end{figure}

On the other hand, the translational correlation functions
obtained for systems with and without random pinning are
qualitatively different (see Fig.~\ref{fig:fig3}). In this case
the densities corresponding to the loss of solid phase stability
are considerably different. It can be deduced from
Fig.~\ref{fig:fig3} (a) that without pinning the stability limit
corresponds to $\rho\approx 0.63$ while in systems having the
quenched disorder the solid-hexatic stability limit is
$\rho\approx 0.66$ (see Fig.~\ref{fig:fig3} (b)).

Fig.~\ref{fig:fig4} depicts the equation of state  at $T=0.2$. It
is seen that the liquid-hexatic stability limit density is located
inside the Van der Waals loop for both systems the one having the
random pinning and without it. At the same time, in pinning-free
systems the hexatic-solid stability limit is located outside the
Van der Waals loop. One can conclude that in this case the system
melts according to the melting scenario proposed in
Refs.~\cite{foh1,foh2,foh5}.

In presence of random pinning the hexatic-solid instability
density shifts to the higher densities, yet, the melting scenario
remains the same. In our case we observe the first-order
liquid-hexatic transition and continuous hexatic-solid transition.
Fig.~\ref{fig:fig5} provides for corresponding behavior in
orientational $\psi_6$ and translational $\psi_T$  order
parameters. The orientational order parameter $\psi_6$ is typical
for conventional behavior, for example, for systems without random
pinning \cite{dfrt2}. However, the translational order parameter
$\psi_T$ has a qualitatively different form. It can be seen in
Fig.~\ref{fig:fig5} that the curve $\psi_T$ has a step-like
behavior. When density decreases, $\psi_T$ moves down at
$\rho\approx 0.66$ corresponding to the solid phase stability
limit in presence of random pinning. On the other hand, $\psi_T$
is not equal to zero at $\rho\approx 0.66$ because the local
translational order exists in this state (see Fig.~\ref{fig:fig3}
and discussion given in previous Section). The translational order
parameter disappears only in the two-phase region.

Since the translational order parameter $\psi_T$ is not equal to
zero in hexatic phase, it is interesting to calculate the
diffusion coefficient in order to be sure that the system is in
the liquid state (having quasi-long range orientational order). It
should be noted that in cases when the hexatic phase was reported
(see, for example, \cite{foh1,foh2,foh5,prest1}) the density range
is extremely narrow, and it is very difficult to calculate the
dynamic properties of the system. In presence of quenched disorder
the hexatic phase gets considerably wider, and the calculation of
diffusion coefficient gives more accurate results. In
Fig.~\ref{fig:fig6} we plot the diffusion coefficient for
$\sigma=1.15$ having random pinning and without it. One can see
that in case without pinning an increase in diffusion coefficient
begins at $\rho\approx 0.63$, while in a system having random
pinning the diffusion coefficient has non-zero values at
$\rho\approx 0.66$ in accordance with results shown in Figs.
\ref{fig:fig4} and \ref{fig:fig5}.

\section{Conclusions}

In this paper we present results of computer simulations of
melting transitions in $2D$ core softened systems having small
length of the repulsive shoulder (Eq. (\ref{3}) and
Fig.~\ref{fig:fig1}) for cases when the random pinning (quenched
disorder) is present and when there is no random pinning. It is
shown that without quenched disorder the system having small
repulsive shoulder $\sigma=1.15$ which is close in shape to soft
disks $1/r^n$ with $n=14$ melts according to the melting scenario
proposed in Refs.~\cite{foh1,foh2,foh4} (first-order
liquid-hexatic and continuous hexatic-solid transitions). Random
pinning widens the hexatic phase without altering the melting
scenario. It is shown that as one can expect the random pinning
almost does not change the behavior of the orientational order
parameter, orientational correlation function and equation of
state of the system. This means that the first order transition
between the hexatic phase and isotropic liquid is almost
unchanged. On the other hand, random pinning drastically changes
the behavior of the translational correlation function
(Fig.~\ref{fig:fig3}) and translational order parameter (see
Fig.~\ref{fig:fig5}). As it was discussed above, this change is
related to the different behavior of the local and overall
translational order in the system. The solid-hexatic transition
shifts to the range of the higher densities. We also calculate the
diffusion coefficient in the hexatic phase (Fig.~\ref{fig:fig6}).
One can see that at the point of the BKT type solid-hexatic
transition diffusion coefficient becomes nonzero and slowly grows
while the diffusion coefficient of the solid phase (without
pinning) at the same densities is zero. The rapid increase of the
diffusion coefficient begins at the two phase region. As one can
expect, the diffusion coefficient of the  system with pinning is
slightly lower than the one in the pinning-free isotropic liquid.
It should be noted that the signs of the solid-hexatic BKT-type
transition cannot be found on the equation of state plot
(Fig.~\ref{fig:fig4}). As it was mentioned above, the melting
criterion based on the behavior of the translational correlation
function gives only the stability limit of the solid phase. In
this case it is important to have additional proofs of the melting
transition. The behavior of the translational order parameter and
diffusion coefficient shown in Figs.~\ref{fig:fig5} and
\ref{fig:fig6} confirms the system melting scenario.

It is interesting to  note that as it was discussed in our
previous publication \cite{dfrt5}, melting in system having larger
repulsive step $\sigma=1.35$ is more complex. At high densities
the conventional first-order transition takes place without random
pinning. It seems that the repulsive shoulder added to the soft
disk potential $1/r^{14}$ makes the hexatic phase which does exist
in soft disks with $n>6$ \cite{foh4} unstable. The disorder,
however, drastically changes this melting scenario. A single
first-order transition is split into two transitions, one of them
(solid-hexatic) is the continuous BKT-like transition, and the
hexatic to isotropic liquid transition occurs as the first order
transition in accordance with \cite{foh1,foh2,foh4}. A possible
mechanism for this transition can be associated with spontaneous
proliferation of grain boundaries \cite{chui83,foh3,foh5}. It
should be noted, that melting scenario with single first-order
transition corresponds to systems kept at sufficiently low
temperatures. At high temperatures the repulsive shoulder of the
potential becomes ineffective, and the properties of the potential
will be similar to that of soft disks $1/r^n$ with $n=14$. In this
case we can expect finding a critical temperature (similar to
tricritical point) above which melting should occur through two
transitions in accordance with scenario proposed in Ref.
\cite{foh4}.

It should be also noted, that the nature of the first-order
liquid-hexatic transition is not completely understood as
conventional theories like the BKTHNY are not capable of
describing the first-order liquid-hexatic transition.

\bigskip

The authors are grateful to S.M. Stishov and V.V. Brazhkin for
valuable discussions. We are thankful to the Russian Science Center at the
Kurchatov Institute and the Joint Supercomputing Center of the Russian
Academy of Science for providing computational facilities. The work was
supported by the Russian Science Foundation (Grant No
14-12-00820).

\end{document}